\documentclass[prb,preprint,showpacs,preprintnumbers,amsmath,amssymb]{revtex4}
\usepackage{graphicx}
\usepackage{dcolumn}
\usepackage{bm}

\begin{document}

\title{Ultrafast electron dynamics and cubic optical nonlinearity of
free standing thin film of double walled carbon nanotubes }

\author{N. Kamaraju$^{1,2}$, Sunil Kumar$^{1,2}$, B. Karthikeyan$^{1,2}$, Alexander Moravsky$^{3}$, R. O. Loutfy$^{3}$ and A.K. Sood$^{1,2}$\footnote{Electronic mail:~asood@physics.iisc.ernet.in}}
\affiliation{$^{1}$Center for Ultrafast Laser Applications(CULA),
Indian Institute of Science,~Bangalore - 560 012, India}
\affiliation{$^{2}$Department of Physics, Indian Institute of
Science,Bangalore - 560 012, India} \affiliation{$^{3}$Materials
and Electrochemical Research Corporation, Tucson, AZ, USA}

\date{\today}

\begin{abstract}
Ultrafast degenerate pump-probe experiments performed on a free
standing film of double walled carbon nanotubes near the first
metallic transition energy of outer tube show ultrafast ($97~fs$)
photobleaching followed by a photo-induced absorption with a slow
relaxation of $1.8~ps$. Femtosecond closed and open aperture
z-scan experiments carried out at the same excitation energy show
saturation absorption and negative cubic nonlinearity. From these
measurements,~real and imaginary part of the third order nonlinear
susceptibility are estimated to be Re($\chi^{(3)}) \sim -2.2
\times 10 ^{-9}~esu$ and Im($\chi^{(3)}) \sim 1.1 \times 10
^{-11}~esu$.

\end{abstract}


 \maketitle
    Hollow cylinders made by folding graphene sheet named as Carbon nanotubes (CNTs),
have attracted intensive research due to their fascinating
physical properties with possible applications such as in
mechanics,\cite{Baughman} flow transducers,\cite{sood1}
nanoelectronics\cite{Collins, Misewich}, and nonlinear
optics.\cite{nonlinear1,nonlinear2,nonlinear3}One dimensional
nature of delocalized~$\pi$-electron cloud along the tube
axis,\cite{Margulis,Lauret_APL} make CNTs the most promising
material with large and ultrafast electronic third order
susceptibility\cite{Maeda} to be utilized in exciting applications
like optical terahertz (THz) switching and passive modelocking.
Femtosecond time resolved photoinduced studies on single walled
nanotubes (SWNT) \cite{Korovyanko, Ichida, Manzoni,Ellingson,
Hertel-adv,Ichida_Physica_B,Lauret_PRL,Ojostic_PRL,Lauret_semi}
reveal that the excitons are the primary photo-excitations in
semiconducting-SWNT (s-SWNT) whereas free carriers are
photoexcited in metallic-SWNT (m-SWNT). These studies indicate
that the photoinduced response changes from fast ($\leq$5 ps) to
slow (30 ps) when the SWNTs are dispersed in a liquid from their
bundled state. The relaxation mechanisms are different in metallic
and semiconducting nanotubes. Electron-electron and
electron-phonon interactions contribute to fast and slow
components, respectively, in m-SWNT \cite{Ichida,Hertel1,Hertel2}.
On the other hand, in s-SWNT intra-band carrier relaxation and
inter-band carrier recombination lead to fast and slow relaxations
respectively\cite{Ichida,Lauret_PRL,Ojostic_PRL,Lauret_semi}.
Environmental effects and defects also play a major role in these
relaxation processes.\cite{Lauret,Ichida} Chen \emph{et
al.}\cite{YCchen}, measured the Im($\chi^{(3)})$ $\sim$ $10^{-10}~
esu$ whereas Maeda \emph{et al.}\cite{Maeda} reported that
Im($\chi^{(3)})$ could be as high as $10^{-6}$ esu in the case of
s-SWNT films on a substrate. Recent z-scan measurements
\cite{kamaraju} at 1.57 eV showed Im($\chi^{(3)})$ = $10^{-9}~esu$
and Re($\chi^{(3)})$ = $-4.4 \times 10^{-9}~esu$ for
SWNT-suspensions (using a surfactant) and Im($\chi^{(3)})$ = $8
\times 10^{-9}~esu$ and Re($\chi^{(3)})$ = $1.4 \times
10^{-8}~esu$ at 0.85 eV for SWNT thin films\cite{recent_APL}. In
the case of multi walled carbon nanotubes (MWNT) grown on a quartz
substrate\cite{Elim}, the values are Im($\chi^{(3)})= -1.6 \times
10^{-11}~esu$ and Re($\chi^{(3)}) = -1.7 \times 10^{-11}~esu$, and
photobleaching decay time of $\sim~2~ps$ was reported. As compared
to SWNT and MWNT, double walled carbon nanotubes (DWNT) have not
been investigated for their nonlinear properties, except the work
of Nakamura \emph{et al.},\cite{Nakamura} where pump-probe studies
show photo bleaching with biexponential relaxation for DWNT
suspensions. The shortest decay time (400 fs) of photobleaching
was attributed to carrier relaxation between inner and outer tubes
and the slow component (4 ps) was ascribed to the recombination of
electrons and holes at the bottom of the bands. Except this study,
there has been no other study to our knowledge on DWNT. In this
letter, we report degenerate Pump-Probe~(PP) measurements and
femtosecond z-scan experiments (both closed aperture (CA) and open
aperture (OA))~on unsupported film of DWNT. We show that the
Im($\chi^{(3)})$ is two orders of magnitude smaller compared to
that of SWNT aqueous suspensions\cite{kamaraju} whereas the
Re($\chi^3$) is of the same order of magnitude.
\begin{figure}[htbp]
\includegraphics[width=0.5\textwidth]{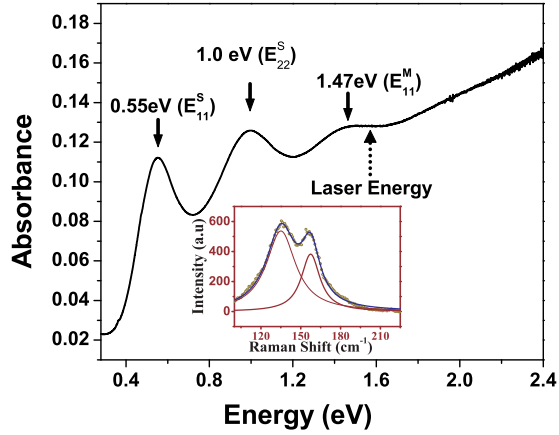}
\caption{Optical absorption spectrum of double walled carbon
nanotube film. $E_{11}^{S}$ and $E_{22}^{S}$  are the first and
second interband transition energies of the semiconducting
nanotubes and $E_{11}^{M}$ is the first transition energy of
metallic tubes. Dotted arrow shows the energy where degenerate PP
and z-scan measurements were done. Inset shows RBM modes at 135
$cm^{-1}$ and 157 $cm^{-1}$ corresponding to the average diameter
of 1.97 and 1.65 nm respectively of the outer tubes.}
\end{figure}

DWNT films of thickness 200 nm (measured using Atomic Force
Microscopy (AFM))~were prepared as reported
elsewhere\cite{Gadagkar} and characterized by Raman spectroscopy~
(using Ar ion laser at 2.41 eV) and optical absorption (Bruker
FT-IR) spectroscopy. Inset of Fig. 1 shows radial breathing modes
(RBM) of DWNT at $135~cm^{-1}$ and $157~cm^{-1}$, corresponding to
the outer tube average diameters of $1.97~nm$ and $1.65~nm$,
respectively. Fig. 1 displays the optical absorption spectrum of
the DWNT film showing three bands at $0.55~eV,~1.0~eV$ and
$1.47~eV$ corresponding to $E_{11}^{S}$, $E_{22}^{S}$ and
$E_{11}^{M}$ electronic transitions of the outer tube of diameter
1.97 nm, as suggested by the theoretical calculations.\cite{Reich}
The dotted arrow in Fig. 1 marks the laser photon energy,~$E_{L}$
of 1.57 eV used in all our experiments which is very close to
$E_{11}^{M}$. The refractive index, $n_{0}$ of the DWNT film is
3.6 at 1.57 eV as calculated from its absorption and transmission
measurements using Fresnel's relation between reflection
coefficient and extinction coefficient.\cite{Pankove}

The output from Ti:Sapphire Regenerative femtosecond amplifier (50
fs, 1.57 eV, 1 KHz Spitfire, Spectra Physics) was used for both
z-scan and the degenerate PP experiments.  At the sample point,
the cross-correlation of pump and probe pulses was measured to be
$75~fs$ (FWHM) using a thin BBO crystal. The pump pulse was
delayed in time using the computer controlled motorized
translation stage (XPS Motion controller, Newport). The change in
the probe transmission due to the presence of the pump was
monitored using two Si-PIN diodes~(one for the reference beam and
the other for the probe beam interacting with the pump) with the
standard lock-in detection~(pump beam was chopped at 139 Hz). The
probe intensity was kept at $58~MW/cm^{2}$ in all the experiments
and the pump intensity was maintained at $1.5
~GW/cm^{2},~1.0~GW/cm^{2},~556~MW/cm^{2}~$and$~303~MW/cm^{2}$ for
intensity dependent PP studies. All these measurements were
performed with pump and probe polarizations perpendicular to each
other to avoid coherent artifacts\cite{Okomoto}. In the
femtosecond z-scan experiments, the intensity was varied from
$150~MW/cm^2$ to $9.4~GW/cm^2$.The pulse from the amplifier was
found to be broadened to 80 fs near the sample point in z-scan
experiments. For the CA z-scan, an aperture of 1~mm size was kept
in front of the detector whereas for OA z-scan, all the
transmitted light was collected.

\begin{figure}[htbp]
   \includegraphics[width=0.5\textwidth]{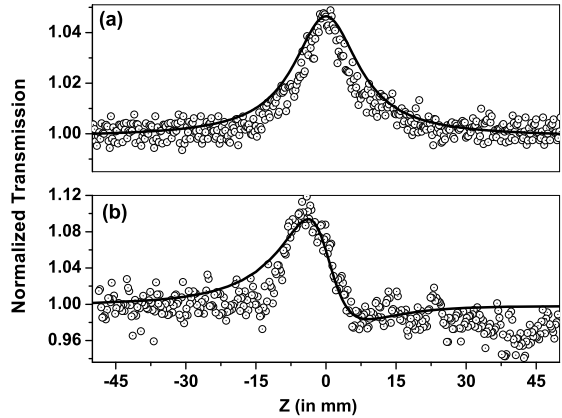}
  \caption{~Normalized transmittance data (open circles) in (a) OA
  z-scan and (b) CA z-scan (aperture linear transmittance = 0.12).
  Theoretical fit (solid line) is obtained with $\beta_{0} = 1.4 \times 10^{-8}~cm/W$, $I_s= 13~GW/cm^{2}$
  and $\gamma \sim -2.6 \times 10^{-11}~cm^2/W$.}
\end{figure}

The femtosecond OA and CA z-scan experiments performed at the same
excitation energy (1.57 eV) reveal saturation absorption and
negative nonlinearity as shown in Fig. 2 by open circles. Using
our modified\cite{kamaraju} approach to analyze z-scan data, both
the OA and CA z-scans were fitted~(shown as solid lines in Fig. 2)
with a consistent set of parameters, $\gamma = -2.6 \times
10^{-11}~cm^2/W,~\beta_{0} = 1.4 \times 10^{-8}~cm/W$ and $I_s =
13~GW/cm^2$. Here, $\gamma$ is the nonlinear refraction
coefficient, $\beta_{0}$ is the two photon absorption (TPA)
coefficient and $I_s$ is the saturation intensity. The third order
nonlinear susceptibility was calculated using\cite{Ganeev}
Im($\chi^{(3)})$=$\frac{6.6\times
10^{-23}c^{2}n_{0}^{2}~\beta_{0}(in~cm/W)}{96~\pi^{2}~E_{L}}$ and
Re($\chi^{(3)})$=$\frac{10^{-7}c~n_{0}^{2}~\gamma(in~cm^{2}/W)}{480~\pi^{2}}$,~
where c is the velocity of the light in vacuum. The measured
values are Im($\chi^{(3)}) = 1.1 \times 10 ^{-11}~esu$ and
Re($\chi^{(3)}) = -2.2 \times 10 ^{-9}~esu$. It can be seen that
as compared to SWNT suspensions\cite{kamaraju}, Im($\chi^{(3)})$
is two orders of magnitude smaller whereas the Re($\chi^{(3)}$) is
of the same order of magnitude.

Degenerate PP data using four different pump intensities is shown
in Fig. 3. The initial decay corresponds to photobleaching (PB)
and the latter part corresponds to photo-induced absorption~(PA).
The inset of Fig. 3 shows PA part of differential transmission
data (normalized to its PB peak height) for three values of the
pump intensities, $303~MW/cm^{2}, 556~MW/cm^{2}$ and $1.0~
GW/cm^{2}$. The data for the pump intensity of $556~ MW/cm^{2}$
overlaps with that of $303~MW/cm^{2}$ and hence is not shown. The
time-dependent photo absorption is fitted with a monoexponential
function (shown as solid line in the inset) with a time constant
of 1.8~ps for the pump intensities $303~MW/cm^{2}$ and
$1.0~GW/cm^{2}$. It can be seen that the PA is negligible for pump
intensity of ~$1.5~GW/cm^{2}$ and was not analyzed. On the other
hand, PB can be resolved for pump intensity of $1.5~ GW/cm^{2}$.
The PB data for this pump intensity is fitted with the convolution
of the cross-correlation of the pump and probe pulses and a
monoexponential function. This yields a time constant of 97 fs.
For the lower pump intensities, the decay time decreases below our
time resolution (75 fs).

\begin{figure}[htbp]
   \includegraphics[width=0.5\textwidth]{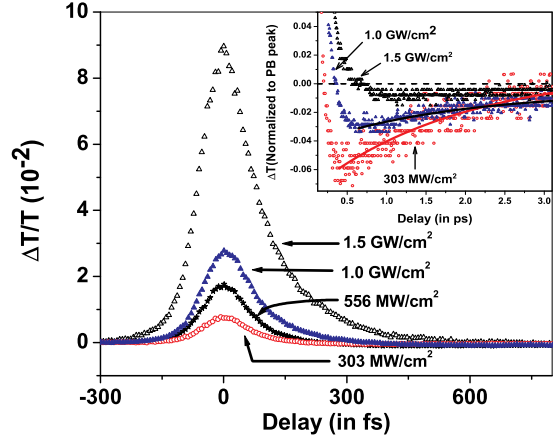}
  \caption{(colour online) Pump intensity dependent $\Delta T/T$ where the probe intensity is kept at $58~MW/cm^{2}$.
   The inset shows PA part of differential transmission
data (normalized to its PB peak height) for three values of the
pump intensities, $303~MW/cm^{2},~556~MW/cm^{2}$ and $1.0~
GW/cm^{2}$. It can be seen from the inset that the PA is almost
negligible for the highest pump intensity. Solid lines are the
single exponential decay fit to the data with the time constant of
1.8 ps.}
\end{figure}

We now offer a plausible explanation of our results. The PB in our
experiments at 1.57 eV can only arise from $E_{11}^{M}$ of m-SWNT
since pump and probe energies are nearly resonant with
$E_{11}^{M}$. The long PA decay time suggests that the excited
state absorption could be happening from $E_{11}^{S}$ excitonic
level and not from $E_{22}^{S}$ excitonic level as earlier PP
studies\cite{Ellingson,Korovyanko} show that PA from $E_{22}^{S}$
is very fast~($\sim 200 fs$). Therefore, the PA will dominantly
arise from s-SWNT~(transition from excitonic level $E_{11}^{S}$ to
$E_{33}^{S}$). Further, PA at 1.57 eV can also arise from m-SWNT
(transition from $E_{11}^{M}$ to $E_{33}^{M}$). At higher
intensities, larger number of charge carriers are produced which
will destroy the excitons in semiconducting tubes. This can result
in the decrease of the PA associated with the s-SWNT.

We have also estimated $Im (\chi^{(3)}$) from PP studies as
follows\cite{Maeda}.
\begin{equation}
Im(\chi^{(3)}) = \frac {\varepsilon_{0} c^{2} n_{0}^{2}}{3 \omega
I_{pump}~\Delta \alpha_{0} (\tau=0)}
\end{equation}
where \begin{equation} \Delta \alpha_{0} (\tau=0) ~= ~\frac
{-\alpha_{pump}~ln~[\frac {\Delta T}{T}(\tau=0) +
1]}{1-e^{-\alpha_{pump}L_{s}}}
\end{equation}
Here, $\alpha_{pump}$ is the linear absorption coefficient of the
sample for the pump beam, $\Delta \alpha_{0}(t)$ is the absorption
change of the probe beam when the pump and probe have zero
delay~($\tau = 0$), $L_{s}$ is the sample thickness,
$\varepsilon_{0}$ is the permittivity of vacuum, $\lambda$ is the
wavelength of the light and I$_{pump}$ is the pump intensity.
Using measured $\alpha_{pump}$ = 6405 $cm^{-1}$ and $L_{s}$ = 200
nm, we get $\beta_{0} = 3.9 \times 10^{-8}$ cm/W which correspond
to Im($\chi^{(3)}$) =~$ 2.1\times 10^{-11}~esu$. This value is
found to be independent of the pump intensity and agrees well with
the value obtained from our z-scan studies. The ultrafast
saturable absorption and large nonlinear third order nonlinearity
means that the unsupported film of DWNTs can be used as saturable
absorbers in the passive optical regeneration, mode-locking and
THz optical switching.

AKS thanks Department of Science and Technology, India, for
support.

\end{document}